****** Start of file apssamp.tex ******
\documentclass[preprint,aps]{revtex4}

\usepackage{graphicx}
\usepackage{dcolumn}
\usepackage{bm}

\begin{document}

 \begin{flushright}
 SLAC--PUB--10774   \\
 \vspace*{-3mm}
 October 2004
 \end{flushright}


\begin{center}
\begin{large}
  {\bf Direct Measurements of $A_b$ and $A_c$ using Vertex/Kaon Charge Tags at SLD$^*$}
\end{large}
\end{center}

%
%
%
%
\begin{center}
\def\iAOMORI{$^{(1)}$}
\def\iBRI{$^{(2)}$}
\def\iBRUN{$^{(3)}$}
\def\iBU{$^{(4)}$}
\def\iCOLO{$^{(5)}$}
\def\iCSU{$^{(6)}$}
\def\iFERR{$^{(7)}$}
\def\iFRAS{$^{(8)}$}
\def\iJHU{$^{(9)}$}
\def\iLBL{$^{(10)}$}
\def\iMASS{$^{(11)}$}
\def\iMISSI{$^{(12)}$}
\def\iMIT{$^{(13)}$}
\def\iMOSCOW{$^{(14)}$}
\def\iNAGO{$^{(15)}$}
\def\iOREG{$^{(16)}$}
\def\iOXF{$^{(17)}$}
\def\iPERU{$^{(18)}$}
\def\iQMU{$^{(19)}$}
\def\iRUTG{$^{(20)}$}
\def\iRAL{$^{(21)}$}
\def\iSLAC{$^{(22)}$}
\def\iSOONG{$^{(23)}$}
\def\iTENN{$^{(24)}$}
\def\iTOHO{$^{(25)}$}
\def\iUCSB{$^{(26)}$}
\def\iUCSC{$^{(27)}$}
\def\iVAND{$^{(28)}$}
\def\iWASH{$^{(29)}$}
\def\iWISC{$^{(30)}$}
\def\iYALE{$^{(31)}$}

   \baselineskip=.75\baselineskip
\mbox{Koya Abe\unskip,\iTOHO}
\mbox{Kenji Abe\unskip,\iNAGO}
\mbox{T. Abe\unskip,\iSLAC}
\mbox{I. Adam\unskip,\iSLAC}
\mbox{H. Akimoto\unskip,\iSLAC}
\mbox{D. Aston\unskip,\iSLAC}
\mbox{K.G. Baird\unskip,\iMASS}
\mbox{C. Baltay\unskip,\iYALE}
\mbox{H.R. Band\unskip,\iWISC}
\mbox{T.L. Barklow\unskip,\iSLAC}
\mbox{J.M. Bauer\unskip,\iMISSI}
\mbox{G. Bellodi\unskip,\iOXF}
\mbox{R. Berger\unskip,\iSLAC}
\mbox{G. Blaylock\unskip,\iMASS}
\mbox{J.R. Bogart\unskip,\iSLAC}
\mbox{G.R. Bower\unskip,\iSLAC}
\mbox{J.E. Brau\unskip,\iOREG}
\mbox{M. Breidenbach\unskip,\iSLAC}
\mbox{W.M. Bugg\unskip,\iTENN}
\mbox{D. Burke\unskip,\iSLAC}
\mbox{T.H. Burnett\unskip,\iWASH}
\mbox{P.N. Burrows\unskip,\iQMU}
\mbox{A. Calcaterra\unskip,\iFRAS}
\mbox{R. Cassell\unskip,\iSLAC}
\mbox{A. Chou\unskip,\iSLAC}
\mbox{H.O. Cohn\unskip,\iTENN}
\mbox{J.A. Coller\unskip,\iBU}
\mbox{M.R. Convery\unskip,\iSLAC}
\mbox{V. Cook\unskip,\iWASH}
\mbox{R.F. Cowan\unskip,\iMIT}
\mbox{G. Crawford\unskip,\iSLAC}
\mbox{C.J.S. Damerell\unskip,\iRAL}
\mbox{M. Daoudi\unskip,\iSLAC}
\mbox{S. Dasu\unskip,\iWISC}
\mbox{N. de Groot\unskip,\iBRI}
\mbox{R. de Sangro\unskip,\iFRAS}
\mbox{D.N. Dong\unskip,\iMIT}
\mbox{M. Doser\unskip,\iSLAC}
\mbox{R. Dubois\unskip,\iSLAC}
\mbox{I. Erofeeva\unskip,\iMOSCOW}
\mbox{V. Eschenburg\unskip,\iMISSI}
\mbox{E. Etzion\unskip,\iWISC}
\mbox{S. Fahey\unskip,\iCOLO}
\mbox{D. Falciai\unskip,\iFRAS}
\mbox{J.P. Fernandez\unskip,\iUCSC}
\mbox{K. Flood\unskip,\iMASS}
\mbox{R. Frey\unskip,\iOREG}
\mbox{E.L. Hart\unskip,\iTENN}
\mbox{K. Hasuko\unskip,\iTOHO}
\mbox{S.S. Hertzbach\unskip,\iMASS}
\mbox{M.E. Huffer\unskip,\iSLAC}
\mbox{X. Huynh\unskip,\iSLAC}
\mbox{M. Iwasaki\unskip,\iOREG}
\mbox{D.J. Jackson\unskip,\iRAL}
\mbox{P. Jacques\unskip,\iRUTG}
\mbox{J.A. Jaros\unskip,\iSLAC}
\mbox{Z.Y. Jiang\unskip,\iSLAC}
\mbox{A.S. Johnson\unskip,\iSLAC}
\mbox{J.R. Johnson\unskip,\iWISC}
\mbox{R. Kajikawa\unskip,\iNAGO}
\mbox{M. Kalelkar\unskip,\iRUTG}
\mbox{H.J. Kang\unskip,\iRUTG}
\mbox{R.R. Kofler\unskip,\iMASS}
\mbox{R.S. Kroeger\unskip,\iMISSI}
\mbox{M. Langston\unskip,\iOREG}
\mbox{D.W.G. Leith\unskip,\iSLAC}
\mbox{V. Lia\unskip,\iMIT}
\mbox{C. Lin\unskip,\iMASS}
\mbox{G. Mancinelli\unskip,\iRUTG}
\mbox{S. Manly\unskip,\iYALE}
\mbox{G. Mantovani\unskip,\iPERU}
\mbox{T.W. Markiewicz\unskip,\iSLAC}
\mbox{T. Maruyama\unskip,\iSLAC}
\mbox{A.K. McKemey\unskip,\iBRUN}
\mbox{R. Messner\unskip,\iSLAC}
\mbox{K.C. Moffeit\unskip,\iSLAC}
\mbox{T.B. Moore\unskip,\iYALE}
\mbox{M. Morii\unskip,\iSLAC}
\mbox{D. Muller\unskip,\iSLAC}
\mbox{V. Murzin\unskip,\iMOSCOW}
\mbox{S. Narita\unskip,\iTOHO}
\mbox{U. Nauenberg\unskip,\iCOLO}
\mbox{H. Neal\unskip,\iYALE}
\mbox{G. Nesom\unskip,\iOXF}
\mbox{N. Oishi\unskip,\iNAGO}
\mbox{D. Onoprienko\unskip,\iTENN}
\mbox{L.S. Osborne\unskip,\iMIT}
\mbox{R.S. Panvini\unskip,\iVAND}
\mbox{C.H. Park\unskip,\iSOONG}
\mbox{I. Peruzzi\unskip,\iFRAS}
\mbox{M. Piccolo\unskip,\iFRAS}
\mbox{L. Piemontese\unskip,\iFERR}
\mbox{R.J. Plano\unskip,\iRUTG}
\mbox{R. Prepost\unskip,\iWISC}
\mbox{C.Y. Prescott\unskip,\iSLAC}
\mbox{B.N. Ratcliff\unskip,\iSLAC}
\mbox{J. Reidy\unskip,\iMISSI}
\mbox{P.L. Reinertsen\unskip,\iUCSC}
\mbox{L.S. Rochester\unskip,\iSLAC}
\mbox{P.C. Rowson\unskip,\iSLAC}
\mbox{J.J. Russell\unskip,\iSLAC}
\mbox{O.H. Saxton\unskip,\iSLAC}
\mbox{T. Schalk\unskip,\iUCSC}
\mbox{B.A. Schumm\unskip,\iUCSC}
\mbox{J. Schwiening\unskip,\iSLAC}
\mbox{V.V. Serbo\unskip,\iSLAC}
\mbox{G. Shapiro\unskip,\iLBL}
\mbox{N.B. Sinev\unskip,\iOREG}
\mbox{J.A. Snyder\unskip,\iYALE}
\mbox{H. Staengle\unskip,\iCSU}
\mbox{A. Stahl\unskip,\iSLAC}
\mbox{P. Stamer\unskip,\iRUTG}
\mbox{H. Steiner\unskip,\iLBL}
\mbox{D. Su\unskip,\iSLAC}
\mbox{F. Suekane\unskip,\iTOHO}
\mbox{A. Sugiyama\unskip,\iNAGO}
\mbox{A. Suzuki\unskip,\iNAGO}
\mbox{M. Swartz\unskip,\iJHU}
\mbox{F.E. Taylor\unskip,\iMIT}
\mbox{J. Thom\unskip,\iSLAC}
\mbox{E. Torrence\unskip,\iMIT}
\mbox{T. Usher\unskip,\iSLAC}
\mbox{J. Va'vra\unskip,\iSLAC}
\mbox{R. Verdier\unskip,\iMIT}
\mbox{D.L. Wagner\unskip,\iCOLO}
\mbox{A.P. Waite\unskip,\iSLAC}
\mbox{S. Walston\unskip,\iOREG}
\mbox{A.W. Weidemann\unskip,\iTENN}
\mbox{E.R. Weiss\unskip,\iWASH}
\mbox{J.S. Whitaker\unskip,\iBU}
\mbox{S.H. Williams\unskip,\iSLAC}
\mbox{S. Willocq\unskip,\iMASS}
\mbox{R.J. Wilson\unskip,\iCSU}
\mbox{W.J. Wisniewski\unskip,\iSLAC}
\mbox{J.L. Wittlin\unskip,\iMASS}
\mbox{M. Woods\unskip,\iSLAC}
\mbox{T.R. Wright\unskip,\iWISC}
\mbox{R.K. Yamamoto\unskip,\iMIT}
\mbox{J. Yashima\unskip,\iTOHO}
\mbox{S.J. Yellin\unskip,\iUCSB}
\mbox{C.C. Young\unskip,\iSLAC}
\mbox{H. Yuta\unskip.\iAOMORI}

\it
   \vskip \baselineskip                   
   \baselineskip=.75\baselineskip   
\iAOMORI
   Aomori University, Aomori, 030 Japan, \break
\iBRI
   University of Bristol, Bristol, United Kingdom, \break
\iBRUN
   Brunel University, Uxbridge, Middlesex, UB8 3PH United Kingdom, \break
\iBU
   Boston University, Boston, Massachusetts 02215, \break
\iCOLO
   University of Colorado, Boulder, Colorado 80309, \break
\iCSU
   Colorado State University, Ft. Collins, Colorado 80523, \break
\iFERR
   INFN Sezione di Ferrara and Universita di Ferrara, I-44100 Ferrara, 
Italy,
\break
\iFRAS
   INFN Laboratori Nazionali di Frascati, I-00044 Frascati, Italy, \break
\iJHU
   Johns Hopkins University,  Baltimore, Maryland 21218-2686, \break
\iLBL
   Lawrence Berkeley Laboratory, University of California, Berkeley, 
California
94720, \break
\iMASS
   University of Massachusetts, Amherst, Massachusetts 01003, \break
\iMISSI
   University of Mississippi, University, Mississippi 38677, \break
\iMIT
   Massachusetts Institute of Technology, Cambridge, Massachusetts 
02139, \break
\iMOSCOW
   Institute of Nuclear Physics, Moscow State University, 119899 Moscow, 
Russia,
\break
\iNAGO
   Nagoya University, Chikusa-ku, Nagoya, 464 Japan, \break
\iOREG
   University of Oregon, Eugene, Oregon 97403, \break
\iOXF
   Oxford University, Oxford, OX1 3RH, United Kingdom, \break
\iPERU
   INFN Sezione di Perugia and Universita di Perugia, I-06100 Perugia, 
Italy,
\break
\iQMU
   Queen Mary, University of London, London, E1 4NS United Kingdom,
\break
\iRUTG
   Rutgers University, Piscataway, New Jersey 08855, \break
\iRAL
   Rutherford Appleton Laboratory, Chilton, Didcot, Oxon OX11 0QX United 
Kingdom,
\break
\iSLAC
   Stanford Linear Accelerator Center, Stanford University, Stanford, 
California
94309, \break
\iSOONG
   Soongsil University, Seoul, Korea 156-743, \break
\iTENN
   University of Tennessee, Knoxville, Tennessee 37996, \break
\iTOHO
   Tohoku University, Sendai, 980 Japan, \break
\iUCSB
   University of California at Santa Barbara, Santa Barbara, California 
93106,
\break
\iUCSC
   University of California at Santa Cruz, Santa Cruz, California 95064, 
\break
\iVAND
   Vanderbilt University, Nashville,Tennessee 37235, \break
\iWASH
   University of Washington, Seattle, Washington 98105, \break
\iWISC
   University of Wisconsin, Madison,Wisconsin 53706, \break
\iYALE
   Yale University, New Haven, Connecticut 06511. \break

\rm
%

\end{center}


  

\begin{center}
{\bf Abstract}
\end{center}

    Exploiting the manipulation of the SLC electron-beam polarization,
    we present precise direct measurements of the parity violation
    parameters $A_c$ and $A_b$ in the $Z$ boson--$c$ quark and
    $Z$ boson--$b$ quark coupling. Quark/antiquark discrimination
    is accomplished via a unique algorithm that takes advantage of the
    precise SLD CCD vertex detector, employing the net charge of
    displaced vertices as well as the charge of kaons that emanate
    from those vertices. From the 1996-98 sample of 400,000 $Z$
    decays, produced with an average beam polarization of 73.4\%, we
    find $A_c = 0.673 \pm 0.029 ({\rm stat.}) \pm 0.023 ({\rm syst.})$ 
    and $A_b = 0.919 \pm 0.018 ({\rm stat.}) \pm 0.017 ({\rm syst.}).$


\begin{center}
{\it Submitted to Physical Review Letters}
\end{center}

\vspace*{15mm}
$^*$ Work supported in part by Department of Energy contract
DE-AC02-76SF00515

\vfill
\eject


Measurements of fermion production asymmetries at the $Z^0$ pole
determine the extent of parity violation in the $Zf\bar{f}$ coupling.
At Born level, the differential cross section for
the process $e^+e^-\rightarrow Z^0 \rightarrow f\bar{f}$
can be expressed
as a function of the polar angle
$\theta$ of the fermion
relative to the electron beam direction,
\begin{equation}
\label{DIFFXSECT}
{d\sigma_f \over d\cos\theta}
 \propto (1 - A_e P_e)(1 + \cos^2 \theta) +
2 A_f (A_e - P_e) \cos \theta ,\nonumber
\end{equation}
where
$P_e$ is the longitudinal polarization of the electron beam  ($P_e > 0$ for
predominantly right-handed polarized beam).
The parameter $A_f = 2v_f a_f/(v_f^2+a_f^2)$, where
$v_f\ (a_f)$ is the vector (axial vector) coupling of the fermion $f$
to the $Z^0$ boson,
expresses the extent of parity violation in the $Zf\bar{f}$ coupling.

From the conventional forward-backward asymmetries formed with an
unpolarized electron beam ($P_e = 0$), such as that used by the CERN
Large Electron-Positron Collider (LEP) experiments, only the product
$A_eA_f$ of parity-violation parameters can be
measured~\cite{p:lewwg99s}.  With a longitudinally polarized electron
beam, however, it is possible to measure $A_f$ independently of $A_e$
by fitting simultaneously to the differential cross sections of
Eq. (1) formed separately for predominantly left- and right-handed
beam. The resulting direct measurement of $A_f$ is largely independent
of propagator effects that modify the effective weak mixing angle,
and thus is complementary to other electroweak
asymmetry measurements performed at the $Z^0$ pole.

In this Letter, we present measurements of $A_c$ and $A_b$ based on the
use of the invariant mass of displaced vertices
to select $Z \rightarrow c {\bar c}$ and $Z \rightarrow b {\bar b}$
events. The charge of the underlying quark is determined via a unique
algorithm that exploits the net charge of the displaced vertices, as
well as the charge of tracks emanating from the vertices that are
identified as kaons.

The operation of the SLC with a polarized electron beam has been described
elsewhere~\cite{p:SLCpol}. During the 1996-98 run, 
the SLC Large Detector (SLD)~\cite{p:SLD,p:anr}
recorded an integrated luminosity of 14.0 pb$^{-1}$, at a mean center-of-mass
energy of 91.24 GeV, and with a luminosity-weighted mean electron-beam polarization
of $|P_e| = 0.734 \pm 0.004$~\cite{p:sldalr}.

The SLD measures charged particle tracks with the Central Drift
Chamber (CDC), which is immersed in a uniform axial magnetic field of
0.6T.  The VXD3 vertex detector provides an accurate measure of
particle trajectories close to the beam axis.  For the 1996-98 data,
the combined $r\phi$ ($rz$) impact parameter resolution of the CDC and
VXD3 is 7.8 (9.7) $\mu$m at high momentum, and 34 (34) $\mu$m at
$p_{\perp}\sqrt{\sin \theta}$ = 1 GeV/c, where $p_{\perp}$ is the
momentum transverse to the beam direction, and $r$ ($z$) is the
coordinate perpendicular (parallel) to the beam axis.  The combined
momentum resolution in the plane perpendicular to the beam axis is $
\delta p_{\bot} / p_{\bot} =
\sqrt{(.01)^2+(.0026~p_{\bot}/GeV/c)^2 \,}$.
A Cerenkov Ring-Imaging Detector (CRID)~\cite{p:crid}, using a
combination of liquid and gaseous radiators, allows efficient $K-\pi$
separation in the range $0.3 \; GeV/c < p_K < \; 30 GeV/c$
for tracks with $|\cos\theta| < 0.68$.
The thrust axis is reconstructed using the Liquid Argon Calorimeter,
which covers the angular range $|\cos \theta| < 0.98$.  We employ a
Monte Carlo (MC) simulation of the production and detection processes
that makes use of the JETSET 7.4 event generator~\cite{p:jetset}, the
QQ~\cite{p:cleob} package for $B$ hadron decay specially tuned to
match the CLEO inclusive $D$ production distributions~\cite{p:cleod}
and the ARGUS particle production distributions~\cite{p:argusd},
and the GEANT 3.21
framework~\cite{p:geant} for the simulation of the SLD detector.

Events are classified as hadronic $Z^0$ decays if they: (1) contain at
least seven well-measured tracks (as described in Ref.~\cite{p:SLD});
(2) exhibit a visible charged energy of at least 18 GeV; (3) have a
thrust axis polar angle satisfying $|\cos\theta_{thrust}| < 0.7$; and
(4) have a thrust magnitude greater than 0.8 (to suppress events with 
both heavy hadrons in the same hemisphere).
Vertex identification is done using a topological algorithm
\cite{p:djnim}, enhanced via the application of a neural-network 
selection based on the flight distance and angle of the reconstructed
vertex\cite{p:thesis}. 
According to the MC, secondary vertices are found in 72.7\% of
bottom-quark, 28.2\% of charm-quark, and 0.41\% of light-quark event hemispheres.
 
Due to the cascade nature of $B$ decays, tracks from the decay may not
all originate from the same space point. An independent neural network,
exploiting the location of the point of closest approach of the track 
to the line connecting the primary and secondary
vertices~\cite{p:thesis}, is used to attach tracks with two or more VXD
hits that are not already included in the secondary vertex. `VXD-only' tracks 
with three or more VXD hits, but no CDC segment, are also considered for
attachment; if attached, the fit vertex location is used as an additional
space point to improve the charge determination.

\begin{figure}
  \caption{\label{f:selnn} Output distribution from the
  flavor-selection neural network; the separate bottom, charm,
  and uds contributions are derived from MC simulation.}
  \centerline{\includegraphics[width=.45\textwidth]{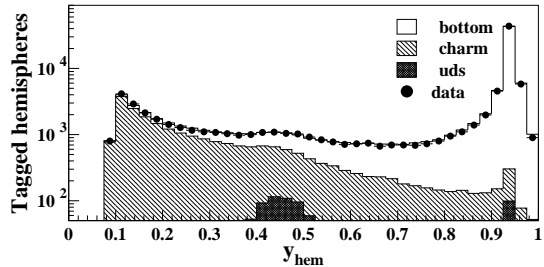}}
  \end{figure}

A final neural network, making use of 
the $p_T$-corrected vertex mass ($M_{VTX}$)~\cite{p:ptcorm},
the total momentum of the vertexed tracks ($P_{VTX}$),
the flight distance from the IP to the vertex, and the
number of tracks in the vertex~\cite{p:thesis}, 
is used to discriminate between 
bottom and charm events. The output $y_{hem}$ of this neural net is 
shown in Figure~\ref{f:selnn}.

The analysis makes use of two mutually exclusive tags. The $L$-tag,
optimized to select $c$ hemispheres, requires $y_{hem}<0.4$ and 
$P_{VTX}>5$ GeV/$c$. The $H$-tag, optimized to select $b$
requires $y_{hem}>0.85$ and $M_{VTX}<7$ GeV/$c^2$. From the
MC simulation, we find that 84\% (98\%) of events with one (two) 
$L$-tagged hemisphere(s), and no $H$-tagged hemispheres, are
$Z \rightarrow c {\bar c}$ decays, while 97\% (100\%) of events 
with one (two)
$H$-tagged hemisphere(s) are $Z \rightarrow b {\bar b}$ decays.

\begin{figure} \caption{\label{f:chgdist} Distributions of
  hemisphere charge: (a) $Q_{vtx}$, $y_{hem}<0.4$; (b) $Q_K$,
  $y_{hem}<0.4$ (the large $Q_K = 0$ contribution is suppressed);
  (c) $Q_{vtx}$, $y_{hem}>0.85$, including VXD-only
  tracks. The $D^+$,$D^-$,$D^0$ and $B^+$,$B^-$,$B^0$
  designations refer to all positive, negative, or neutral heavy
  flavor hadrons, including baryons, from the MC simulation.}
  \centerline{\includegraphics[width=.45\textwidth]{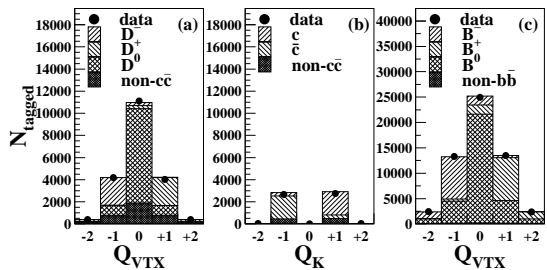}}
  \end{figure}

Within tagged hemispheres, two quantities are used to discriminate
quark from antiquark production: the net charge of all vertexed tracks
($Q_{VTX}$) and the net charge of all vertex tracks that are
identified as kaons ($Q_K$).  The presence of a quark is indicated by
$Q_{VTX} > 0$ or $Q_K < 0$ for the $L$-tag, and $Q_{VTX} < 0$ for the
$H$-tag; for this latter tag, the kaons do not make a significant
additional contribution.  If an $L$ or $H$-tagged hemisphere cannot be
assigned a nonzero charge using these methods, or if an $L$-tag has
both $Q_{VTX}$ and $Q_K$ nonzero and in disagreement, it is treated as
untagged. The resulting charge distributions are shown in
Figure~\ref{f:chgdist}.


By comparing the tagging and sign-determination results between
hemispheres in data events, it is possible to determine most of the per-hemisphere
tagging efficiencies $\epsilon_f^T$ and their correct-sign
probabilities $p_f^T$ for the tags $T=L,H$\cite{p:thesis}. The
fractions of single-$H$ ($XH$), double-$H$ ($HH$), mixed ($HL$),
single-$L$ ($XL$) and double-$L$ ($LL$) tagged events are sensitive to
the hemisphere tagging efficiencies; a maximum likelihood fit to these
fractions is used to constrain the values of $\epsilon_c^L$,
$\epsilon_b^L$, $\epsilon_c^H$, and $\epsilon_b^H$
(Table~\ref{t:perhemi}), assuming Standard Model (SM) values 
for the fraction $R_b$ ($R_c$) of $b {\overline b}$
($c {\overline c}$) events from $e^+e^-$ annihilation at the
$Z^0$ pole. 
The hemisphere correct-sign probabilities
$p_c^L$, $p_b^L$, and $p_b^H$ (Table~\ref{t:perhemi}) are constrained by a
similar fit to the fractions of oppositely-signed hemispheres in $HH$,
$HL$, and $LL$ events, using the previously determined tagging
efficiencies as input.

In constraining the tagging efficiencies and correct-sign
probabilities from the inter-hemisphere tagging and signing
information, it is necessary to account for inter-hemisphere
correlations that alter the nominal relationship between
single-hemisphere and full-event tagging and charge signing
performance. MC studies confirm that, for vertex-based tagging and
signing, inter-hemisphere correlations are
due primarily to correlation in the energy and angle of the hadrons
containing the heavy quarks, and from events for which both heavy
hadrons are produced in the same hemisphere. To account for these
effects, we have used the MC simulation to explore the dependence of
the tagging and signing parameters as a function of the number of
heavy hadrons in the hemisphere, and of the polar angle and energy
(after restricting to hemispheres with a single heavy hadron) of the
heavy hadron. The effects of the inter-hemisphere correlations can
then be accounted for by convolving these dependences with the
distributions of the number, energy, and polar angle of heavy hadrons
within and opposite to tagged hemispheres, as described
in~\cite{p:thesis}.  Ignoring these effects would
incorrectly lower $A_b$
and $A_c$ by approximately 1.5\% of their subsequent fit values.

  \begin{table}
    \caption{\label{t:perhemi} Per-hemisphere
    efficiencies (requiring that hemispheres be tagged
    and have non-zero net charge) $\epsilon_v^T$ and
    correct-sign probabilities $p_f^T$.
    ``Calib'' refers to the values
    obtained from the calibration procedure described
    in~\cite{p:thesis}, while the ``MC'' column shows the expectations
    from the simulation.  $p_c^H$ is not calibrated from the data and so
    is not shown.}
    \begin{tabular}{c|cc|cc}
       & $\epsilon_{MC}$ & $\epsilon_{calib}$ & $p_{MC}$ & $p_{calib}$ \\
      \hline
      $c$, $L$-tag & 0.121 & 0.115$\pm$0.002 & 0.932 & 0.918$\pm$0.010 \\
      $b$, $L$-tag & 0.020 & 0.022$\pm$0.001 & 0.545 & 0.543$\pm$0.031 \\
      $c$, $H$-tag & 0.005 & 0.006$\pm$0.002 & --- & --- \\
      $b$, $H$-tag & 0.323 & 0.325$\pm$0.002 & 0.807 & 0.821$\pm$0.005
    \end{tabular}
  \end{table}

The fit for the parameter $A_c$ makes use of events with at least one
$L$-tagged hemisphere and neither $H$-tagged, while the $A_b$ fit uses
events with at least one H tag.  Events with two L or two H tags are
discarded if the charges in the two hemispheres are in disagreement.
For events with one H and one L tag, only the H tag is used to sign
the thrust axis. From the MC simulation, we find that
84\% (98\%) of events with one L tag (two L tags) are $Z \rightarrow c \bar{c}$ 
events, while
97\% (100\%) of events with one H tag (two H tags) are $Z \rightarrow b \bar{b}$
events.

Unbinned maximum-likelihood fits are performed to the Born-level 
differential cross section:
  \begin{equation}
    {\cal L} \sim
    (1-A_eP_e)(1+\cos^2\theta_{\hat{t}})+2(A_e-P_e)A_E\cos\theta_{\hat{t}}
  \end{equation}

\noindent where $\theta_{\hat{t}}$ is the polar angle of the
signed thrust axis, and the electron beam polarization is signed so
that $P_e > 0$ for right-handed electrons.  The fitted effective asymmetry
$A_E$ is given by the sum over the flavor composition of the sample:
  \begin{equation}
    A_E = \sum_f \Pi_f(2P_f-1)(1-C_f^{QCD})(A_f-\delta_f^{QED})
  \end{equation}

\noindent where
$\Pi_f$ is the fraction of and $P_f$ the correct-signing probability
for the flavor $f$, calculated separately for single- and double-tagged
events, making use of the values in
Table~\ref{t:perhemi} when possible. For the light flavor $uds$
contribution, the simulated mistag rates are used for $\Pi_{uds}$,
while $P_{uds}$ is set to $0.5 \pm 0.29$ (uniform probability between
0 and 1). Mechanisms for developing a charm signal in the $H$-tagged
sample tend to favor incorrect charge assignment~\cite{p:thesis},
leading to the assumption $p_c^H = 0.25 \pm 0.14$.  Because events at
larger values of $|\cos\theta_{\hat{t}}|$ carry larger statistical
weights in the fits, but poorer overall tagging qualities, the MC
simulation is used to parameterize these values as a function of
$\cos\theta_{\hat{t}}$.  Failing to account for this effect would
incorrectly lower the fitted values of $A_c$ and $A_b$ by 1-2\%.

The corrections $C_f^{QCD}$ for gluon radiation are evaluated as
in~\cite{p:lepqcd}.  The ${\cal O}(\alpha_S^2)$
corrections are evaluated in~\cite{p:lewwg99s} as 4.5\%(3.8\%) for
$c$($b$) events, using the calculation in~\cite{p:csqcd} based on the
parton thrust axis (we ignore the hadronization corrections 
of~\cite{p:lewwg99s} since they are implicit in our signed thrust axis
analyzing power). Additionally, the analysis procedure suppresses
events with hard gluon radiation, and so these results are further
scaled by factors $s_f$ of 0.27$\pm$0.13(0.53$\pm$0.08) for $C_c^{QCD}$
($C_b^{QCD}$), as determined by the MC simulation.

The $\delta_f^{QED}$ terms correct the asymmetries for the effects of
initial-state QED radiation and $\gamma/Z$ interference, and are
determined by ZFITTER~\cite{p:zfitter} to be $\delta_c^{QED} = 0.0012$
and $\delta_b^{QED} = -0.0021$.



From a sample of 9970 events, using the SM value
$A_b = 0.935$ as input, we obtain
$A_c = 0.6747 \pm 0.0290 (\rm{stat.})$, while
from a sample of 25917 events, using the Standard Model value of
$A_c = 0.667$ as input, we obtain
$A_b = 0.9173 \pm 0.0184 (\rm{stat.})$.

  

We have explored a number of potential sources of systematic error;
these are summarized in 
Table II. 
For both $A_c$ and
$A_b$, the dominant systematic uncertainty arises from the limited
statistics available for the calibration of the purity of the
flavor-selected sample, and of the correct-sign tagging probability of
the sample, resulting in a relative systematic uncertainty
of $\pm 3.0\%$ ($\pm 1.5\%$) for
$A_c$ ($A_b$). 
By studying a sample enriched in $uds$ quark
production ($M_{hem} < 2$ GeV/c$^2$ and $P_{hem} < 4$ GeV/c), the
fake-vertex efficiency $\epsilon_{uds}$ is constrained to be within
25\% of its MC expectation, leading to uncertainties of $\pm 0.1\%$
($\pm 0.0\%$) on $A_c$ ($A_b$).  

The procedure for calibrating the sample purity and correct-signing
probabilities is subject to uncertainties in the correlation between 
the quark and antiquark energies, and in the fraction of events for which 
the quark and antiquark appear in the same hemisphere.
Comparisons between
data and MC of the correlation between the heavy hadron energies in $c$-quark 
and $b$-quark enriched samples constrain the $c$ and $b$ hadron energy
correlations to be within 2.6\% and 0.3\% of their MC expectation (Table II),
while comparisons of samples enhanced in three-jet production showed
the same-hemisphere production rates to be within 1.1\% and 0.7\% of
their MC expectations. The resulting overall
uncertainty in $A_c$ ($A_b$) due
to tagging correlations is found to be 0.6\% (0.3\%).  

The correction
coefficients $C_f^{theory}$ for hard gluon radiation (`QCD
corrections') are subject to uncertainties in $\alpha_S$, quark
masses, and missing higher order terms, given by~\cite{p:lewwg99s} as $\pm
0.0063$ for both $f=c,b$. The determination of the scale factor $s_f$
applied to account for the selection bias against events with hard
gluon radiation is limited by Monte Carlo statistics to $\pm 0.13$
($\pm 0.08$) for $s_c$ ($s_b$). The resulting overall uncertainty in
the QCD correction is $\pm 0.6\%$ ($\pm 0.4\%$) for $A_c$ ($A_b$).

Adding all sources of systematic error in quadrature, we find
\begin{eqnarray}
A_c & = & 0.6747 \pm 0.0290 ({\rm stat.}) \pm 0.0233 ({\rm syst.}) \\
A_b & = & 0.9173 \pm 0.0184 ({\rm stat.}) \pm 0.0173 ({\rm syst.}). 
\end{eqnarray}

Averaging these results (V) with complementary results for $A_b$ using
momentum-weighted track charge (Q)~\cite{p:sldjetq} and the charge of identified
kaons from secondary vertices for data prior to 1996 (K)~\cite{p:sldkq},
for $A_c$ using fully-reconstructed charmed-meson decays (D)~\cite{p:sldddstar},
and for $A_c$ and $A_b$ together using identified leptons
(L)~\cite{p:sldleptons2}, we arrive at the
overall SLD average of~\cite{p:BLUE}
\begin{eqnarray}
A_c & = & 0.6712 \pm 0.0224 ({\rm stat.}) \pm 0.0157 ({\rm syst.}) \\
A_b & = & 0.9170 \pm 0.0147 ({\rm stat.}) \pm 0.0145 ({\rm syst.})
\end{eqnarray}
\noindent independent of the extent of parity violation in the coupling of the
electron to the $Z^0$ boson, consistent with the Standard Model
expectations of $A_c = 0.667$ and $A_b = 0.935$. 

Alternatively, $A_b$ and $A_c$ can be extracted from LEP
measurements of
the unpolarized heavy-quark forward-backward asymmetries $A_{FB}^{0,Q}$ 
via the relation $A_{FB}^{0,Q} = {3 \over 4} A_Q A_e$. The 
values~\cite{p:LEPEW04} $A_{FB}^{0,c} = 0.0702 \pm 0.0035$ and 
$A_{FB}^{0,b} = 0.0998 \pm 0.0017$, from fits solely to LEP
data, combined with the value~\cite{p:LEPEW04} 
$A_e = 0.1501 \pm 0.0016$ derived from
leptonic forward-backward and
leptonic polarization asymmetries measured at LEP and SLD,
determine the heavy-quark coupling parity violation
parameters to be $A_c = 0.624 \pm 0.032$ and
$A_b = 0.887 \pm 0.018$, consistent with the direct 
measurements provided by the polarized differential cross-section 
data from SLD.

We thank the staff of the SLAC accelerator department for their
outstanding efforts on our behalf. This work was supported by the
U.S. Department of Energy (in part by Contract No. DE-AC02-76SF00515),
the U.S. National Science Foundation, the UK Particle Physics and
Astronomy Research Council, the Istituto Nazionale di Fisica Nucleare
of Italy and the Japan-US Cooperative Research Project on High Energy
Physics.

  \bibliographystyle{apsrev}
  \bibliography{vtx_chg}

\begin{thebibliography}{19}
\expandafter\ifx\csname natexlab\endcsname\relax\def\natexlab#1{#1}\fi
\expandafter\ifx\csname bibnamefont\endcsname\relax
  \def\bibnamefont#1{#1}\fi
\expandafter\ifx\csname bibfnamefont\endcsname\relax
  \def\bibfnamefont#1{#1}\fi
\expandafter\ifx\csname citenamefont\endcsname\relax
  \def\citenamefont#1{#1}\fi
\expandafter\ifx\csname url\endcsname\relax
  \def\url#1{\texttt{#1}}\fi
\expandafter\ifx\csname urlprefix\endcsname\relax\def\urlprefix{URL }\fi
\providecommand{\bibinfo}[2]{#2}
\providecommand{\eprint}[2][]{\url{#2}}

\bibitem[{\citenamefont{Abbaneo et~al.}(2000)}]{p:lewwg99s}
\bibinfo{author}{\bibfnamefont{D.}~\bibnamefont{Abbaneo}} \bibnamefont{et~al.}
  (\bibinfo{collaboration}{LEP Electroweak Working Group}),
  \bibinfo{journal}{CERN-EP-2000-016}  (\bibinfo{year}{2000}).

\bibitem[{\citenamefont{et~al}(1997)}]{p:SLCpol}
\bibinfo{author}{\bibfnamefont{K.~Abe} \bibnamefont{et~al.}},
  \bibinfo{journal}{Phys. Rev. Lett.} \textbf{\bibinfo{volume}{78}},
  \bibinfo{pages}{2075} (\bibinfo{year}{1997}).

\bibitem[{\citenamefont{et~al}(1996)}]{p:SLD}
\bibinfo{author}{\bibfnamefont{K.~Abe} \bibnamefont{et~al.}},
  \bibinfo{journal}{Phys. Rev.} \textbf{\bibinfo{volume}{D53}},
  \bibinfo{pages}{1023} (\bibinfo{year}{1996}).

\bibitem[{\citenamefont{et~al}(2001)}]{p:anr}
\bibinfo{author}{\bibfnamefont{K.~Abe} \bibnamefont{et~al.}},
  \bibinfo{journal}{Ann. Rev. Nucl. Part. Sci.} \textbf{\bibinfo{volume}{51}},
  \bibinfo{pages}{345} (\bibinfo{year}{2001}).

\bibitem[{\citenamefont{Abe et~al.}(2000)}]{p:sldalr}
\bibinfo{author}{\bibfnamefont{K.}~\bibnamefont{Abe}} \bibnamefont{et~al.}, 
\bibinfo{journal}{Phys. Rev. Lett.}
  \textbf{\bibinfo{volume}{84}}, \bibinfo{pages}{5945} (\bibinfo{year}{2000}),
  \eprint{hep-ex/0004026}.

\bibitem[{\citenamefont{K. Abe et~al.}(1990)}]{p:crid}
\bibinfo{author}{\bibfnamefont{K.}~\bibnamefont{Abe}}
  \bibnamefont{et~al.}, \bibinfo{journal}{Nucl. Instrum. Meth.}
  \textbf{\bibinfo{volume}{A343}}, \bibinfo{pages}{74} (\bibinfo{year}{1994}).

\bibitem[{\citenamefont{Sjostrand}(1994)}]{p:jetset}
\bibinfo{author}{\bibfnamefont{T.}~\bibnamefont{Sjostrand}},
  \bibinfo{journal}{Comput. Phys. Commun.} \textbf{\bibinfo{volume}{82}},
  \bibinfo{pages}{74} (\bibinfo{year}{1994}).

\bibitem[{p:c(unpublished)}]{p:cleob}
\bibinfo{journal}{{QQ} - The {CLEO} Event Generator,
  http://www.lns.cornell.edu/public/CLEO/soft/QQ}
  (\bibinfo{year}{unpublished}).

\bibitem[{\citenamefont{L. Gibbons et~al.}(1997)}]{p:cleod}
\bibinfo{author}{\bibfnamefont{L.}~\bibnamefont{Gibbons}}~\bibnamefont{et~al.},
  \bibinfo{journal}{Phys. Rev.} \textbf{\bibinfo{volume}{D56}},
  \bibinfo{pages}{3783} (\bibinfo{year}{1997}).

\bibitem[{\citenamefont{H. Albrecht et~al.}(1989)}]{p:argusd}
\bibinfo{author}{\bibfnamefont{H.}~\bibnamefont{Albrecht}}~\bibnamefont{et~al.},
  \bibinfo{journal}{Z. Phys.} \textbf{\bibinfo{volume}{C44}},
  \bibinfo{pages}{547} (\bibinfo{year}{1989}).

\bibitem[{\citenamefont{Brun et~al.}(1987)\citenamefont{Brun, Bruyant, Maire,
  McPherson, and Zanarini}}]{p:geant}
\bibinfo{author}{\bibfnamefont{R.}~\bibnamefont{Brun}},
  \bibinfo{author}{\bibfnamefont{F.}~\bibnamefont{Bruyant}},
  \bibinfo{author}{\bibfnamefont{M.}~\bibnamefont{Maire}},
  \bibinfo{author}{\bibfnamefont{A.~C.} \bibnamefont{McPherson}},
  \bibnamefont{and} \bibinfo{author}{\bibfnamefont{P.}~\bibnamefont{Zanarini}},
  \bibinfo{type}{Tech. Rep.} \bibinfo{number}{CERN-DD/EE/84-1}
  (\bibinfo{year}{1987}).

\bibitem[{\citenamefont{Jackson}(1997)}]{p:djnim}
\bibinfo{author}{\bibfnamefont{D.~J.} \bibnamefont{Jackson}},
  \bibinfo{journal}{Nucl. Instrum. Meth.} \textbf{\bibinfo{volume}{A388}},
  \bibinfo{pages}{247} (\bibinfo{year}{1997}).

\bibitem[{\citenamefont{Wright}(2002)}]{p:thesis}
\bibinfo{author}{\bibfnamefont{T.}~\bibnamefont{Wright}},
  \bibinfo{journal}{SLAC-R-602}  (\bibinfo{year}{2002}).

\bibitem[{\citenamefont{Abe et~al.}(1998)}]{p:ptcorm}
\bibinfo{author}{\bibfnamefont{K.}~\bibnamefont{Abe}} \bibnamefont{et~al.},
  \bibinfo{journal}{Phys. Rev. Lett.}
  \textbf{\bibinfo{volume}{80}}, \bibinfo{pages}{660} (\bibinfo{year}{1998}).

\bibitem[{\citenamefont{Abbaneo et~al.}(1998)}]{p:lepqcd}
\bibinfo{author}{\bibfnamefont{D.}~\bibnamefont{Abbaneo}} \bibnamefont{et~al.}
  (\bibinfo{collaboration}{LEP Heavy Flavor Working Group}),
  \bibinfo{journal}{Eur. Phys. J.} \textbf{\bibinfo{volume}{C4}},
  \bibinfo{pages}{185} (\bibinfo{year}{1998}).

\bibitem[{\citenamefont{Catani and Seymour}(1999)}]{p:csqcd}
\bibinfo{author}{\bibfnamefont{S.}~\bibnamefont{Catani}} \bibnamefont{and}
  \bibinfo{author}{\bibfnamefont{M.~H.} \bibnamefont{Seymour}},
  \bibinfo{journal}{JHEP} \textbf{\bibinfo{volume}{07}}, \bibinfo{pages}{023}
  (\bibinfo{year}{1999}), \eprint{hep-ph/9905424}.

\bibitem[{\citenamefont{Bardin et~al.}(2001)}]{p:zfitter}
\bibinfo{author}{\bibfnamefont{D.}~\bibnamefont{Bardin}} \bibnamefont{et~al.},
  \bibinfo{journal}{Comput. Phys. Commun.} \textbf{\bibinfo{volume}{133}},
  \bibinfo{pages}{229} (\bibinfo{year}{2001}), \eprint{hep-ph/9908433}.

\bibitem[{\citenamefont{Abe et~al.}(2003)}]{p:sldjetq}
\bibinfo{author}{\bibfnamefont{K.}~\bibnamefont{Abe}} \bibnamefont{et~al.}, 
\bibinfo{journal}{Phys. Rev. Lett.}
  \textbf{\bibinfo{volume}{90}}, \bibinfo{pages}{141804}
  (\bibinfo{year}{2003}), \eprint{hep-ex/0208044}.

\bibitem[{\citenamefont{Abe et~al.}(1999{\natexlab{a}})}]{p:sldkq}
\bibinfo{author}{\bibfnamefont{K.}~\bibnamefont{Abe}} \bibnamefont{et~al.}, 
\bibinfo{journal}{Phys. Rev. Lett.}
  \textbf{\bibinfo{volume}{83}}, \bibinfo{pages}{1902}
  (\bibinfo{year}{1999}{\natexlab{a}}).

\bibitem[{\citenamefont{Abe et~al.}(2001)}]{p:sldddstar}
\bibinfo{author}{\bibfnamefont{K.}~\bibnamefont{Abe}} \bibnamefont{et~al.}, 
\bibinfo{journal}{Phys. Rev.}
  \textbf{\bibinfo{volume}{D63}}, \bibinfo{pages}{032005}
  (\bibinfo{year}{2001}), \eprint{hep-ex/0009035}.

\bibitem[{\citenamefont{Abe et~al.}(1999{\natexlab{b}})}]{p:sldleptons2}
\bibinfo{author}{\bibfnamefont{K.}~\bibnamefont{Abe}} \bibnamefont{et~al.}, 
\bibinfo{journal}{Phys. Rev. Lett.}
  \textbf{\bibinfo{volume}{83}}, \bibinfo{pages}{3384}
  (\bibinfo{year}{1999}{\natexlab{b}}).

\bibitem[{\citenamefont{$\rm{Statistical}$ $\rm{and}$ systematic correlations
  between the measurements were taken into account via the Best Linear Unbiased
  Estimator ({BLUE}) algorithm described~by L.~Lyons
  et~al.}(1988)\citenamefont{$\rm{Statistical}$ $\rm{and}$ systematic
  correlations between the measurements were taken into account via the Best
  Linear Unbiased Estimator ({BLUE}) algorithm described~by L.~Lyons, Gibaut,
  and Clifford}}]{p:BLUE}
\bibinfo{author}{\bibnamefont{
  Statistical correlation coefficients $\rho_{xy}$ were found to be
  $\rho_{VQ}=0.32$, $\rho_{VL}=0.15$, $\rho_{QL}=0.22$,
  $\rho_{QK}=0.08$, $\rho_{VK}=0.00$,
  $\rho_{LK}=0.04$ for the $A_b$ measurement, and $\rho_{VD}=0.12$,
  $\rho_{VL}=0.05$, $\rho_{DL}=0.07$ for the $A_c$ measurement.
  $\rm{Statistical}$ $\rm{and}$ systematic
  correlations between the measurements were taken into account via the Best
  Linear Unbiased Estimator ({BLUE}) algorithm described~by L.~Lyons. 
 }},
  \bibinfo{author}{\bibfnamefont{D.}~\bibnamefont{Gibaut}}, \bibnamefont{and}
  \bibinfo{author}{\bibfnamefont{P.}~\bibnamefont{Clifford}},
  \bibinfo{journal}{Nucl. Instr. Meth.} \textbf{\bibinfo{volume}{A270}},
  \bibinfo{pages}{110} (\bibinfo{year}{1988}).

\bibitem[{\citenamefont{Abe et~al.}(1999{\natexlab{b}})}]{p:LEPEW04}
\bibinfo{author}{\bibfnamefont{D.}~\bibnamefont{Abbaneo}} \bibnamefont{et~al.}
\bibnamefont{(LEP Electroweak Working Group)},
\eprint{hep-ex/0412015},
  \bibinfo{year}{December, 2004}.

\end{thebibliography}

\vspace{50mm}

TABLE II: Relative
    systematic errors for the $A_c$ and $A_b$ measurements, in percent (\%).
   A `+' (`-') sign indicates that $A_f$ increases (decreases)
   if the true value of
   the parameter is larger than expected.
  Corrections to the Monte Carlo tracking efficiency and resolution
  simulation have been determined from data;
``Remove'' refers to the difference in the result for $A_f$ when the
  corrections are not applied.

  \begin{table}
    \label{t:systematics}
    \centering
    \begin{tabular}{lccc}
      source & variation & $\delta A_c/A_c$ & $\delta A_b/A_b$ \\
      \hline
	  {\bf Calibration statistics} & & \\
	  ~~$P_f$ & data statistics & 2.96 & 1.41 \\
	  ~~$\Pi_f$ & data statistics & 0.68 & 0.63 \\
	  {\bf EW parameters} & & & \\
	  ~~$R_c$ & 0.1723$\pm$0.0031 & -0.18 & +0.07 \\
	  ~~$R_b$ & 0.2163$\pm$0.0007 & +0.25 & -0.24 \\
	  ~~$A_c$ & 0.667$\pm$0.027 & n/a & +0.04 \\
	  ~~$A_b$ & 0.935$\pm$0.021 & -0.06 & n/a\\
	  {\bf Detector modeling} & & & \\
	  ~~tracking efficiency & remove & -0.36 & +0.34 \\
	  ~~tracking resolution & remove & -0.49 & +0.04 \\
	  ~~CRID $\pi$ mis-ID & data $\pm 1\sigma$ & -0.12 & +0.00 \\
	  {\bf QCD correction} & & & \\
	  ~~$C_f^{theory}$ & $\pm$0.0063 & +0.18 & +0.35 \\
	  ~~$s_f$ & $\pm$0.13, $\pm$0.08 & +0.59 & +0.31 \\
	  {\bf Backgrounds} & & & \\
	  ~~$p_c^H$ & 0.25$\pm$0.14 & +0.83 & -0.56 \\
	  ~~$g\rightarrow c\bar{c}$ & 2.96$\pm$0.38\% & +0.22 & +0.01 \\
	  ~~$g\rightarrow b\bar{b}$ & 0.254$\pm$0.051\% & +0.06 & -0.02 \\
	  ~~fake vertex $\epsilon_{uds}$ & $\pm$25\% & +0.13 & -0.01 \\
	  ~~fake vertex $A_{uds}^{raw}$ & $\pm$0.6 & -0.43 & -0.09 \\
 	  {\bf Tagging correlations} & & & \\
	  ~~same-hemisphere $c\bar{c}$ & 2.82$\pm$1.13\% & +0.33 & -0.01 \\
	  ~~same-hemisphere $b\bar{b}$ & 2.45$\pm$0.74\% & -0.04 & +0.21 \\
	  ~~$c$ energy correlation & 1.4$\pm$2.6\% & +0.48 & -0.14 \\
	  ~~$b$ energy correlation & 1.4$\pm$0.3\% & -0.07 & +0.10 \\
	  {\bf Other} & & & \\
	  ~~Beam polarization & $\pm$0.5\% & -0.50 & -0.50 \\
	  ~~MC statistics & $\pm$1$\sigma$ & 0.64 & 0.34 \\
	  \hline
	      {\bf Total} & & 3.48 & 1.89 \\
    \end{tabular}
  \end{table}
%

\end{document}